# Comparison of 35 and 50 μm thin HPK UFSD after neutron irradiation up to $6 \cdot 10^{15}$ neq/cm$^2$

Y. Zhao, N. Cartiglia[1], E. Estrada, Z. Galloway, C. Gee, A. Goto, Z. Luce,
S. M. Mazza[2], F. McKinney-Martinez, R. Rodriguez, H. F.-W. Sadrozinski, A. Seiden
*SCIPP, Univ. of California Santa Cruz, CA 95064, USA*

V. Cindro, G. Kramberger, I. Mandić, M. Mikuž, M. Zavrtanik
*Jožef Stefan institute and Department of Physics, University of Ljubljana, Ljubljana, Slovenia*

*Abstract–* We report results from the testing of 35 μm thick Ultra-Fast Silicon Detectors (UFSD produced by Hamamatsu Photonics (HPK), Japan and the comparison of these new results to data reported before on 50 μm thick UFSD produced by HPK. The 35 μm thick sensors were irradiated with neutrons to fluences of 0, $1 \cdot 10^{14}$, $1 \cdot 10^{15}$, $3 \cdot 10^{15}$, $6 \cdot 10^{15}$ neq/cm$^2$. The sensors were tested pre-irradiation and post-irradiation with minimum ionizing particles (MIPs) from a $^{90}$Sr β-source. The leakage current, capacitance, internal gain and the timing resolution were measured as a function of bias voltage at -20$^o$C and -27$^o$C. The timing resolution was extracted from the time difference with a second calibrated UFSD in coincidence, using the constant fraction method for both. Within the fluence range measured, the advantage of the 35 μm thick UFSD in timing accuracy, bias voltage and power can be established.

## 1 - Introduction

We are developing a new type of silicon detector, the so-called ultra-fast silicon detector (UFSD) that will establish a new paradigm for space-time particle tracking [1]. The UFSD is a single device that ultimately will measure with high precision concurrently the space (~10 μm) and time (~10 ps) coordinates of a particle.
UFSD are thin pixelated n-on-p silicon sensors based on the Low-Gain Avalanche Detector (LGAD) design [2][3][4] developed by the Centro Nacional de Microelectrónica (CNM) Barcelona, in part as a RD50 Common Project [5]. The sensor exhibits moderate internal gain (~5-70) due to a highly doped p+ region just below the n-type implants. In [6] a time resolution below 35 ps was achieved in a beam test with un-irradiated 45 μm thick UFSD fabricated by CNM. This result complemented previous measurements on thicker sensors in beam tests and laboratory reported in [7][8]. These sets of measurements, taken with LGAD of different thicknesses, agreed well with the predictions of the simulation program Weightfield2 (WF2) [9].
First applications of UFSD are envisioned in the upgrades of the ATLAS and CMS experiments at the High-Luminosity Large Hadron Collider (HL-LHC [10]) as reviewed in [11]. In both experiments, the UFSD would be of moderate segmentation (a few mm$^2$) and will face challenging radiation requirements (fluences up to several $10^{15}$ neq/cm$^2$ and several hundred of MRad). Results on irradiated CNM LGAD 300 μm, 75 μm and 45 μm thick sensors were presented in [12][13]. Recently a radiation campaign with neutrons with 50 μm thick LGAD produced by Hamamatsu Photonics, Japan (HPK) has been reported [14]. In all cases the timing resolution has been shown to deteriorate with fluence due to the decreasing value of the gain. This effect is caused by the acceptor removal mechanism [15] that decreases the concentration of the active dopant in the gain layer.

[1] Permanent address: INFN, Torino, Italy
[2] Corresponding author: simazza@ucsc.edu, telephone (831) 459 1293, FAX (831) 459 5777

In this paper, we report on the performances of 35 µm thick UFSD produced by HPK after a neutron irradiation of fluences of 0, $1\cdot10^{14}$, $1\cdot10^{15}$, $3\cdot10^{15}$, $6\cdot10^{15}$ neq/cm$^2$. In Section 2 we will describe the characteristics of 35 µm thick UFSD, followed in Section 3 by a short description of the irradiation facility. In Section 4, a short description of the experimental set-up is presented, details were previously reported in [6][14]. In Section 5, we will describe the data analysis including the extraction of the gain and the time resolution, and in Section 6 the results on bias dependence of charge collection and gain, pulse characteristics and timing resolution for a range of fluences will be presented. In the same section, the performances of the 35 µm and 50 µm thick UFSD (50D) from HPK will be compared.

**2 – Properties of the HPK UFSDs B35 (35 µm thick) and 50D (50 µm thick)**

The B35 device is a 35 µm thick UFSD sensor from HPK, making it the thinnest UFSD tested in an irradiation campaign to date. The sensor is 1x1 mm$^2$, it is not protected by a guard ring and it has a capacitance of 4.6 pF. The behavior was measured with I-V and C-V scans before irradiation and after neutron irradiation with fluences $1\cdot10^{14}$, $1\cdot10^{15}$, $3\cdot10^{15}$, $6\cdot10^{15}$ neq/cm$^2$. Gain and time resolution were determined in the UCSC β-telescope at two temperatures (-20$^o$C and -27$^o$C) and the results were compared to the HPK 50 µm thick UFSD 50D reported in [14].

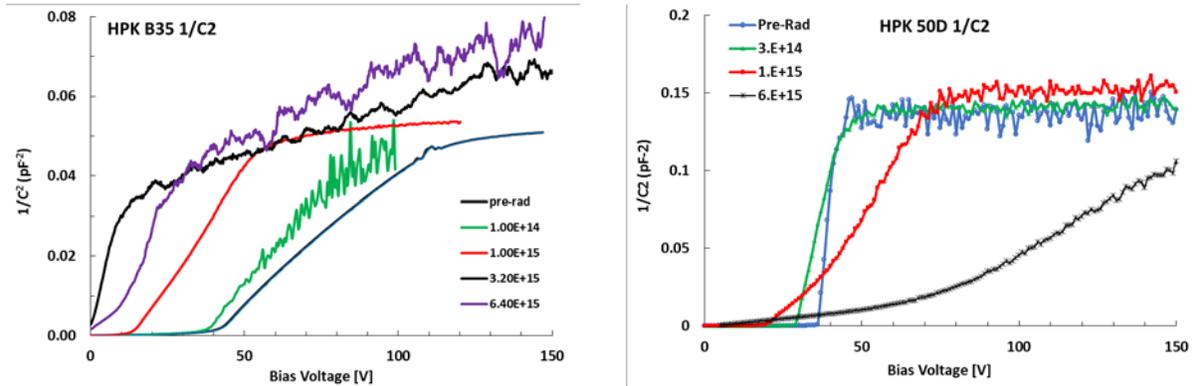

Fig. 1: $1/C^2$ vs. bias voltage for the 35 µm thin B35 (left) and the 50 µm thin 50D (right) (from [14]) at room temperature before irradiation and at -20$^o$C after neutron irradiation to the fluences indicated.

The C-V measurements can be used to extract changes in the doping profile in both the multiplication layer and the bulk. The C-V curves for the two detectors are shown in Fig. 1 where the $1/C^2$ are plotted vs. the bias voltage. The intercept with the bias voltage axis is proportional to the doping density in the multiplication layer, and the slope of the curve by the doping density of the bulk. The effects of acceptor removal in the multiplication layer and the acceptor creation in the bulk are clearly visible by the shortening of the foot (the region at low voltage where the $1/C^2$ curve appears flat before a sudden increase) and the changes in slope.
 For the 50D detector (Fig. 1, right), the slope and the length of the "foot" are monotonically decreasing with fluence. For the B35 detectors (Fig. 1, left) the length of the "foot" is about the same up to $1\cdot10^{14}$ n/cm$^2$, following a large decrease at $1\cdot10^{15}$. The slope is similar between the pre-irradiation case and the three fluences for the interplay of acceptor creation by deep traps and initial acceptor removal by irradiation [1]. The evolution of the doping densities of the multiplication layer and of the bulk for the 50D and B35 sensors are shown in Fig. 2, overlapped to the model of acceptor creation and removal explained in [1].

## 3 – Neutron irradiations

The UFSD were irradiated without bias in the JSI research reactor of TRIGA type in Ljubljana, which has been used successfully in the past decades to support sensor development [17]. The neutron spectrum and flux are well known and the fluence is quoted in 1 MeV equivalent neutrons per cm² (neq/cm² or shortened n/cm²). After irradiation, the devices were annealed for 80 min at 60°C. Afterward the devices were kept in cold storage at -20°C.

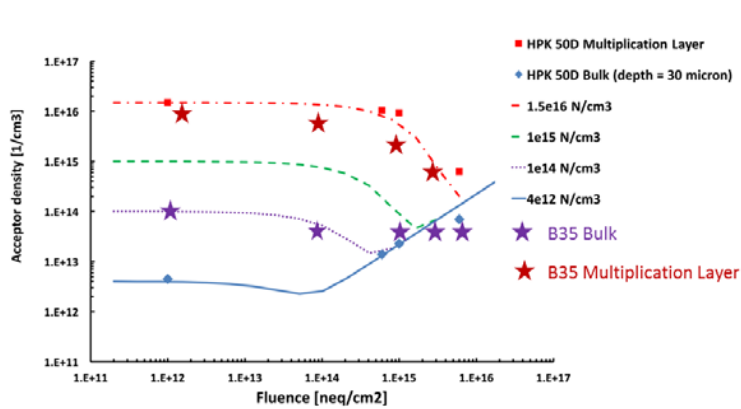

Figure 2. The fluence evolution of the doping concentration in the multiplication layer and the bulk for the 35 μm thick B35 and 50 μm thick 50D (taken from [14]). The curves are simulations described in [1] based on data of [16].

## 4 – β-telescope setup

The laboratory setup with $^{90}$Sr β-source as well as the readout electronics has been previously described in detail in [6],[14]. It is important to note that the system is housed in a climate chamber allowing operations of irradiated sensors at lower temperature down to -30°C. The trigger and time reference is provided by a second UFSD, which in case of the 50D measurements is a CNM UFSD with a time resolution of 27 ps at -20°C, while for the B35 measurements it is a HPK UFSD with time resolution of 15 ± 1 ps. The time resolution for the trigger UFSD was measured pairing two identical UFSDs. Following a trigger, the traces of both trigger and DUT were recorded, with a rate of a few Hz with a digital scope with an analogue bandwidth of 3 GHz and a digitization step of 50 ps.

## 5 – Data analysis

The analysis follows the steps listed in [6]; additional details of the analysis can be found in [14] [18]. The digital oscilloscope records the full voltage waveform of both trigger and DUT in each event, so the complete event information is available for offline analysis. The normalized average pulse shape for both sensors before and after irradiation can be seen in Fig. 3: the tails present in the B35 pre-rad pulses indicate that the bias voltage was not sufficiently high to saturate the drift velocity.

The time of arrival of a particle is defined with the constant fraction discriminator (CFD) method [14] [18], offering a very efficient correction to the time walk effect. The CFD value can be optimized for every bias voltage and fluence to minimize the time resolution, a procedure that is necessary since both the pulse shape and the noise contributions change with fluence. Due to the oscilloscope digitization steps, the time of arrival at a specific CFD fraction is evaluated with a linear interpolation. The event selection is straightforward: for a valid trigger pulse, the signal amplitude, Pmax, of the DUT UFSD should not be saturated by either the scope or the read-out chain. To eliminate the contributions from non-gain events or noise, the time of the pulse maximum, Tmax, has to fall within a window of 1 ns centered on the CFD threshold of 20% of the trigger. The DUT time resolution is calculated from the RMS value of the Gaussian fit to the time difference Δt between the DUT and the trigger.

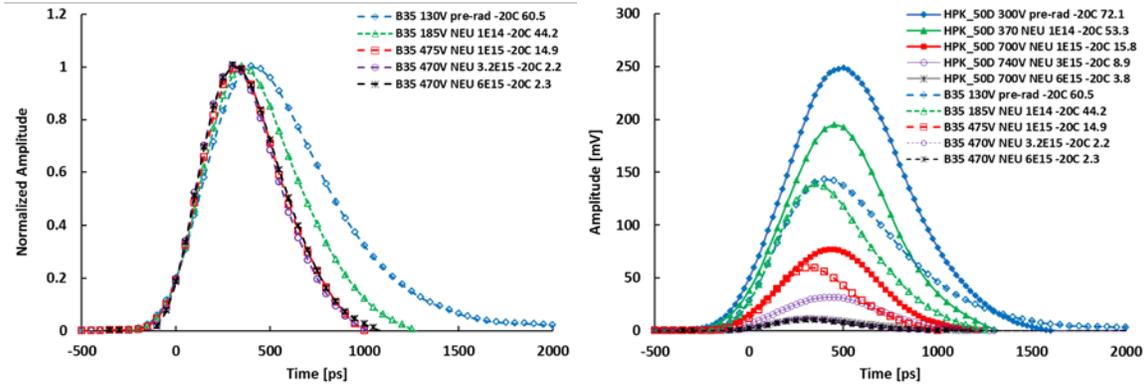

Fig 3: Average normalized pulse shape for B35 (left) and average pulse shapes for B35 and 50D sensors (right). The legends are bias, fluence, temperature, gain.

## 6 - Results on irradiated 35 μm thick UFSD HPK B35

The gain in the sensor is an important parameter for the time resolution. A comparison of the voltage dependence of the gain (Fig. 4, left) shows that B35 requires a lower bias voltage than 50D for the same gain, and that the bias voltage gap between the two sensors of ~ 200V (at the same gain value) is approximately preserved through the radiation steps. The optimum operating voltage is shown in Fig. 4, right. After a fluence of $1 \cdot 10^{15}$ neq/cm$^2$ the bias voltages required are 500 V (B35) and 700 V (50D). The difference in bias and thickness influences the power dissipation. In order to reach a time resolution of 33 ps, 50D will dissipate four times the power than B35.

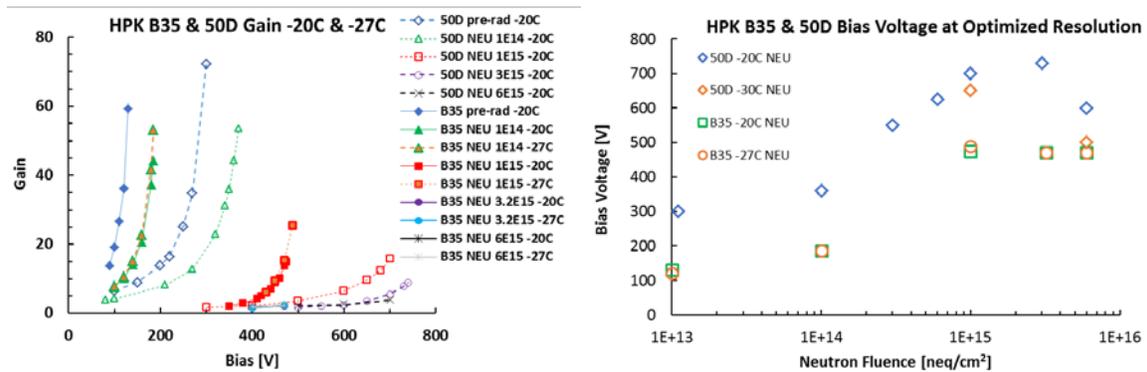

Fig 4: Left: Gain as a function of bias for HPK sensors B35 and 50D. Right: optimum operating voltage for different fluences.

Another parameter determining the time resolution is the rise time. As expected the rise time (Fig. 5) for the 35μm thick detector is significantly shorter than for the 50 μm detector at all fluences. In both sensors the rise time decreases with progressive irradiation.

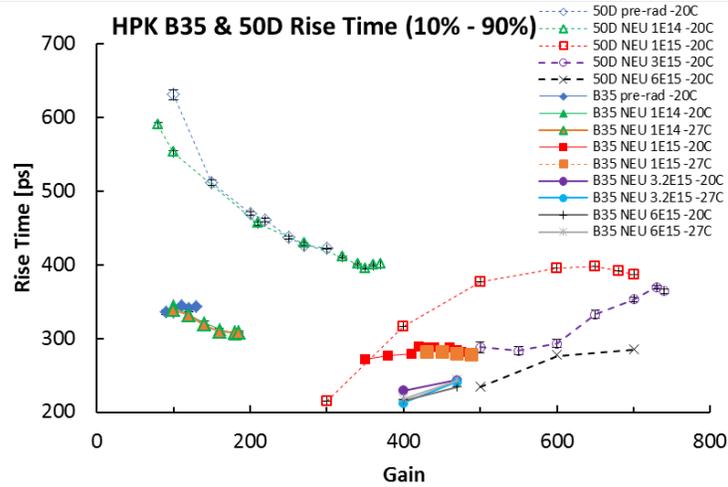

Fig 5: Rise time as a function of bias voltage for sensors B35 (- 20C and -27C) and 50D (- 20C).

Fig. 6 shows the CFD scan of the time resolution for B35 and 50D. The optimal range for the CFD fraction, defined as the range to be within 10% of the optimal time resolution, is for B35 before irradiation: 7-36%, after $1\cdot10^{14}$: 12-90%, after $1\cdot10^{15}$: 15-99%, after $3\cdot10^{15}$: 25-95%, after $6\cdot10^{15}$: 35-95%. These numbers indicate a very flat dependence of the time resolution on the CFD fraction for B35. For 50D the optimal DFC range tends to be narrower.

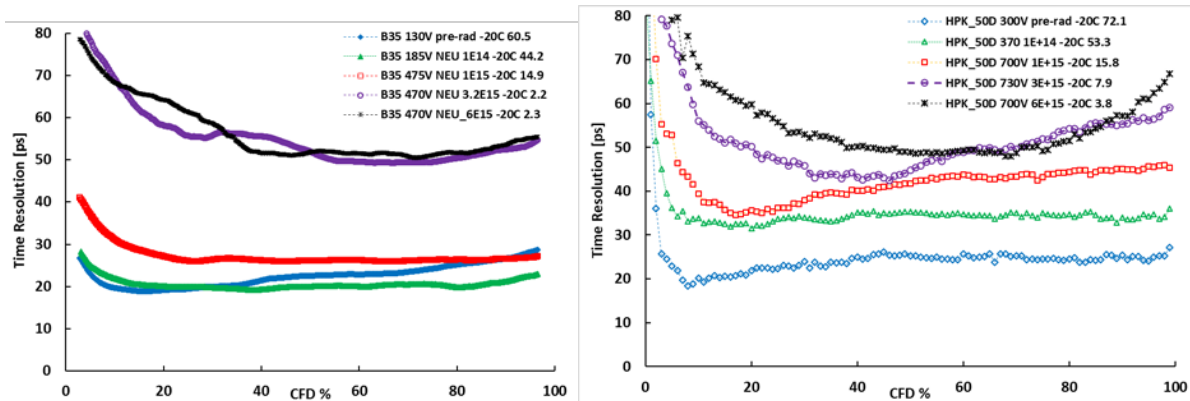

Fig 6: CFD scan at optimum operating voltage for sensors B35 (left) and 50D (right). The legends are bias, fluence, temperature, gain.

The time resolution as a function of gain and bias voltage for different fluences can be found in Fig. 7. It can be seen that the B35 detector requires lower bias, and after irradiation up to $1\cdot10^{15}$ n/cm$^2$ has superior time resolution when compared to 50D. In addition, it exhibits a visible better resolution at -27$^o$C than at -20$^o$C, while the 50D detector show less improvement with lower temperature at the same fluence. The time resolution of the B35 detector at $1\cdot10^{14}$ n/cm$^2$ is better than before irradiation since when new B35 is operated close to depletion.

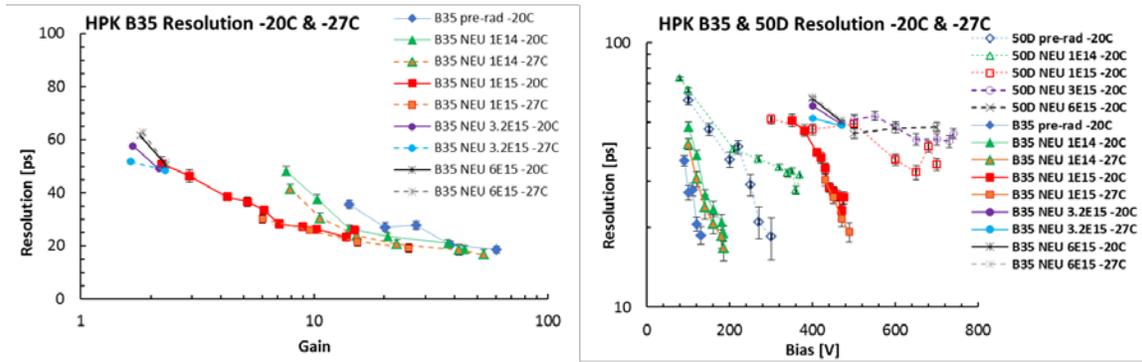

Fig 7: Time resolution as a function of gain for B35 (left) and bias voltage (right) for sensors B35 and 50D. The values are for the CFD optimized for the optimal time resolution.

In Fig.8 the time resolution for B35 and 50D is shown as a function of gain before (left) and after irradiation (right). We expect that at large gain the resolution has a constant value limited by Landau fluctuations, which is smaller for 35 μm than for 50 μm UFSD. This is indeed true for the irradiated sensors on the right, while this fact couldn't be measured before irradiation due to the low bias voltage of B35 which limits the drift velocity of the electrons and holes to sub-optimal values.

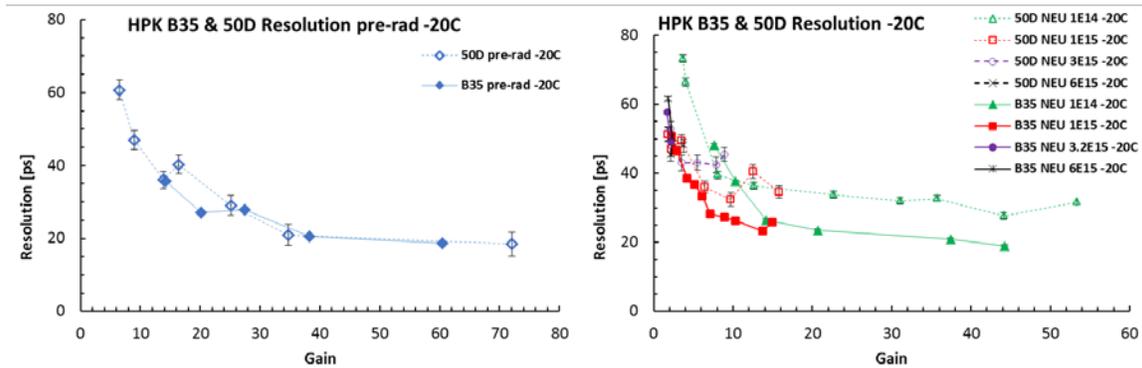

Fig 8: Time resolution as a function of gain for B35 and 50D pre-rad (left) and after irradiation (right).

The time resolutions as a function of fluence are presented in Fig. 9. The optimal CFD fraction used for B35 is 26% before irradiation, 43% after $1\cdot 10^{14}$, 40% after $1\cdot 10^{15}$, 60% after $3\cdot 10^{15}$, and 70% after $6\cdot 10^{15}$ while for 50D it is 8% before irradiation, 20% after $1\cdot 10^{14}$, 20% after $1\cdot 10^{15}$, 48% after $1\cdot 10^{15}$ and 54% after $1\cdot 10^{15}$.

Fig. 9 shows the improvement of the 35 μm thick B35 over the 50 μm thick 50D. Up to a fluence of $1\cdot 10^{15}$ n/cm$^2$ the time resolution is below 25 ps at -20°C, and below 20 ps at -27°C. This is about 10 ps lower than for the 50 μm thick 50D.

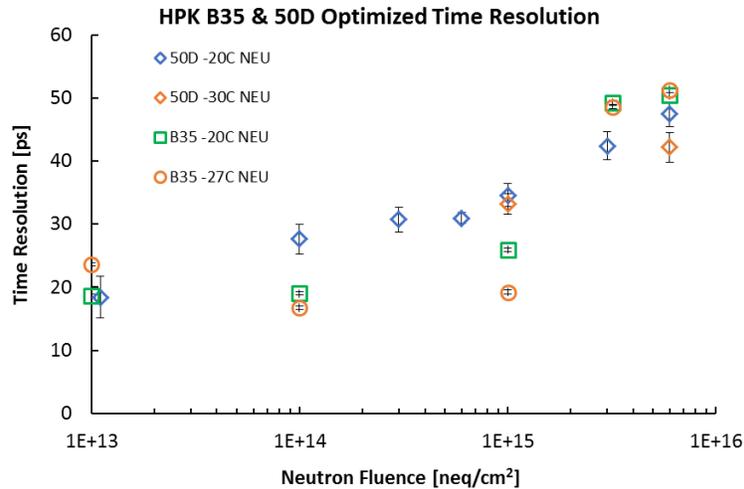

Fig 9: Time resolution as a function of fluence for sensors B35 and 50D.

## 7 - Conclusions

Two UFSD from Hamamatsu Photonics with respective thickness of 35 µm (B35) and 50 µm (50D) were tested using a $^{90}$Sr β-telescope after neutron irradiation up to $1 \cdot 10^{15}$ n/cm$^2$.

The operating bias for the B35 sensor is about 200 V lower than that of the 50D sensor.

The CFD scans of B35 reveal an almost flat dependence of the time resolution on the CFD fraction.
A constant CFD fraction of about 50% will result in a satisfactory time resolution for all fluences

The two detectors show similar time resolution before irradiation: 19 ps for B35 vs. 18 ps for 50D (tested at -20$^o$C). Here the operating bias for B35 is close to depletion.

After a neutron fluence of $1 \cdot 10^{15}$ n/cm$^2$ at -20$^o$C, the time resolution for B35 is 23 ps and for 50D 35 ps, and at -27$^o$C B35 has a time resolution of 19 ps compared with 33 ps for 50D at -30$^o$C.

At higher neutron fluence than $1 \cdot 10^{15}$ n/cm$^2$ B35 has slightly worse resolution than 50D because of lower gain. The performance after large fluences is limited by the breakdown voltage of the UFSD. Thus an important prototyping goal will be increasing the voltage reach of the UFSD.

The B35 sensor shows a dissipated power that is significantly less than the power dissipated by the 50D sensor. After a fluence of $1 \cdot 10^{15}$ n/cm$^2$, to reach a resolution of 33 ps, 50D dissipates 4 times the power than B35.

UFSD of 35 µm thickness are thus good candidates for the use in the ATLAS and CMS upgrade projects for the HL-LHC.

## 8 - Acknowledgements

We acknowledge the contribution to this paper by the HPK team of K. Yamamoto, S. Kamada, A. Ghassemi, K. Yamamura and the expert technical help by the SCIPP technical staff.
Part of this work has been performed within the framework of the CERN RD50 Collaboration.

The work was supported by the United States Department of Energy, grant DE-FG02-04ER41286. Part of this work has been financed by the European Union's Horizon 2020 Research and Innovation funding program, under Grant Agreement no. 654168 (AIDA-2020) and Grant Agreement no. 669529 (ERC UFSD669529), and by the Italian Ministero degli Affari Esteri and INFN Gruppo V.

# 9 - References

[1] H.F.-W. Sadrozinski, A. Seiden and N. Cartiglia, "4D tracking with ultra-fast silicon detectors", 2018 Rep. Prog. Phys. 81 026101.

[2] G. Pellegrini, et al., "Technology developments and first measurements of Low Gain Avalanche Detectors (LGAD) for high energy physics applications", Nucl. Instrum. Meth. A765 (2014) 24.

[3] H.F.-W. Sadrozinski, et al., "Ultra-fast silicon detectors", Nucl. Instrum. Meth. A831 (2016) 18.

[4] M. Carulla et al., "First 50 µm thick LGAD fabrication at CNM", 28th RD50 Workshop, Torino, June 7th 2016, https://agenda.infn.it/getFile.py/access?contribId=20&sessionId=8&resId=0&materialId=slides&confId=11109.

[5] RD50 collaboration, http://rd50.web.cern.ch/rd50/.

[6] N. Cartiglia et al., "Beam test results of a 16 ps timing system based on ultra-fast silicon detectors", Nucl. Instrum. Meth. A850, (2017), 83–88.

[7] N. Cartiglia et al, "Performance of Ultra-Fast Silicon Detectors", JINST 9 (2014) C02001.

[8] N. Cartiglia, "Design optimization of ultra-fast silicon detectors", Nucl. Instrum. Meth. A796 (2015) 141-148.

[9] F. Cenna, et al., "Weightfield2: A fast simulator for silicon and diamond solid state detector", Nucl. Instrum. Meth. A796 (2015) 149; http://personalpages.to.infn.it/~cartigli/Weightfield2/Main.html.

[10] HL-LHC, http://dx.doi.org/10.5170/CERN-2015-005.

[11] L. Gray, "4D Trackers", at "Connecting the dots", Paris 2017, https://indico.cern.ch/event/577003/contributions/2476434/attachments/1422143/2180715/20170306_LindseyGray_CDTWIT.pdf.

[12] G. Kramberger et al.: "Radiation hardness of thin LGAD detectors", TREDI 2017, https://indico.cern.ch/event/587631/contributions/2471705/attachments/1414923/2165831/RadiationHardnessOfThinLGAD.pdf.

[13] J. Lange et al "Gain and time resolution of 45 µm thin LGAD before and after irradiation up to a fluence of $10^{15}$ neq/cm$^2$", JINST **12** P05003

[14] Z. Galloway et al, "Properties of HPK UFSD after neutron irradiation up to 6e15 n/cm$^2$" arXiv:1707.04961.

[15] G. Kramberger et al., "Radiation effects in Low Gain Avalanche Detectors after hadron irradiations", JINST 10 P07006, 2015.

[16] G. Kramberger, et al., "Effective trapping time of electrons and holes in different silicon materials irradiated with neutrons, protons and pions", Nucl. Instrum. Methods Phys. Res. A 481 (2002) 297–305

[17] Snoj, G. ˇZerovnik and A. Trkov, "Computational analysis of irradiation facilities at the JSI TRIGA reactor", Appl. Radiat. Isot. 70 (2012) 483.

[18] Y Zhao, UCSC Senior Thesis 2017,
https://drive.google.com/drive/folders/0ByskYealR9x7bFY1ZS1pZW9SRWs.